\documentclass[a4paper,11pt]{article}
\usepackage{pos}
\usepackage{graphicx}
\usepackage{dcolumn}
\usepackage{bm}
\usepackage{amsmath}
\usepackage{amssymb}
\usepackage{xspace}
\usepackage{lineno}

\newcommand {\meanpT}    {\ensuremath{\langle p_{\mathrm{T}} \kern-0.1em\rangle}\xspace}
\newcommand {\mean}[1]   {\ensuremath{\langle #1 \kern-0.1em\rangle}\xspace} 
\newcommand {\sqrtsNN}   {\ensuremath{\sqrt{s_{\textsc{NN}}}}\xspace}

\newcommand {\MeanNpart} {\mbox{\ensuremath{< \kern-0.15em N_{part} \kern-0.15em >}}}

\newcommand {\mass}     {\mbox{\rm MeV$\kern-0.15em /\kern-0.12em c^2$}}
\newcommand {\tev}      {\mbox{${\rm TeV}$}\xspace}

\newcommand {\mmom}     {\mbox{\rm MeV$\kern-0.15em /\kern-0.12em c$}}
\newcommand {\gmom}     {\mbox{\rm GeV$\kern-0.15em /\kern-0.12em c$}}
\newcommand {\mmass}    {\mbox{\rm MeV$\kern-0.15em /\kern-0.12em c^2$}}
\newcommand {\gmass}    {\mbox{\rm GeV$\kern-0.15em /\kern-0.12em c^2$}}

\newcommand {\dg}       {\mbox{$\kern+0.1em ^\circ$}}

\newcommand{\gevc}{\ensuremath{\mathrm{GeV}/c}\xspace}

\newcommand{\pt}{\ensuremath{p_{\rm T}}\xspace}

\newcommand{\rmLambdas}         {\ensuremath{\mathrm {\Lambda \kern-0.2em + \kern-0.2em \overline{\Lambda}}}\xspace}

\newcommand {\ptreco}        {\ensuremath{p_{\mathrm{T,jet}}^\mathrm{reco,ch}}\xspace}

\newcommand {\ptch}{\ensuremath{p}_{\mathrm{T,jet}}^{\mathrm{ch}}}

\newcommand {\pttrig} {p_{\mathrm{T,trig}}}

\newcommand {\TTsig}{\mathrm{TT_{Sig}}}
\newcommand {\TTref}{\mathrm{TT_{Ref}}}

\title{Jet acoplanarity via hadron+jet measurements in Pb--Pb collisions at $\sqrtsNN = 5.02~\tev$ with ALICE} 
\ShortTitle{Jet acoplanarity via h+jet measurements with ALICE}

\author*[a]{Jaime Norman}

\affiliation[a]{University of Liverpool,\\
  Oliver Lodge Laboratory, Oxford St, Liverpool, L69 7ZE, UK}

\emailAdd{jaime.norman@cern.ch}

\abstract{We present the first fully-corrected semi-inclusive distribution of charged jets recoiling from a trigger hadron in 0--10\% Pb--Pb collisions at $\sqrt{s_{\mathrm{NN}}}$ = 5.02 TeV as a function of the azimuthal angle between the trigger hadron and jet, $\Delta\varphi$. This technique provides a precise data-driven subtraction of the large uncorrelated background contaminating the measurement, and enables the exploration of jet acoplanarity. 
Results for $R=0.2$ recoil jets in the region $30 < \ptch < 40~\gevc$ are shown, where a suppression and narrowing of the $\Delta\varphi$ distribution is observed in 0--10\% Pb--Pb collisions with respect to a PYTHIA pp reference.}

\FullConference{%
  HardProbes2020\\
  1-6 June 2020\\
  Austin, Texas}

\begin{document}
\maketitle

\section{Introduction}

The measurement of jets recoiling from a high-\pt hadron is sensitive to jet azimuthal broadening effects.
In vacuum, broadening effects occur via Sudakov radiation~\cite{Chen:2016vem}.
In medium, additional jet deflection may occur via multiple soft scatterings, resulting in modification of the azimuthal correlation between the trigger hadron and the recoiling jet~\cite{Chen:2016vem,Gyulassy:2018qhr}.
In addition, the tail of this azimuthal correlation is sensitive to Moli\`ere scatterings off quasi-particles in the medium~\cite{DEramo:2018eoy}. A search for these phenemona in Run-1 of the LHC\footnote{The data taking period at the LHC from 2009-2013} data using hadron-jet acoplanarity showed no evidence of jet broadening with respect to the vacuum expectation within experimental uncertainties~\cite{Adam:2015doa}. However, lower recoiling jet $p_\mathrm{T}$ configurations should be more sensitive to in-medium modifications to the acoplanarity~\cite{Chen:2016vem,Gyulassy:2018qhr,DEramo:2018eoy}.
Recent theoretical work~\cite{Zakharov:2020sfx} suggests that radiative corrections to in-medium modification may be negative and comparable to the non-radiative contribution, which could suppress the broadening or even narrow the azimuthal jet distribution with respect to vacuum.
This contribution reports the first exploration of hadron-jet acoplanarity in central (0--10\%) Pb--Pb collisions using high-statistics Run-2 data\footnote{The data taking period from 2015-2018}, with emphasis on the region of low recoil jet $p_\mathrm{T}$, and is the first such analysis that has been fully corrected to the particle level.

The measurement reported here is the trigger-normalised semi-inclusive yield of jets recoiling from a trigger hadron $\frac{1}{N^\mathrm{AA}_\mathrm{trig}} \frac{\mathrm{d^3}N\mathrm{^{AA}_{jet}}}{\mathrm{d}p^{\mathrm{ch}}_\mathrm{T,jet} \mathrm{d}\Delta\varphi \mathrm{d}\eta_\mathrm{jet}} \bigg|_{p_\mathrm{T,trig} \in \mathrm{TT}} $, where $\ptch$ is the jet transverse momentum and $\Delta\varphi$ is the azimuthal angle between the trigger hadron and a reconstructed jet. The observable $\Delta_\mathrm{recoil} $ is then defined as the difference between the trigger-normalised recoil jet distributions in Signal ($\TTsig$) and Reference ($\TTref$) trigger track $\pt$ ($\pttrig$) intervals~\cite{Adam:2015doa}:


\begin{equation}
\Delta_\mathrm{recoil} = \frac{1}{N^\mathrm{AA}_\mathrm{trig}} \frac{\mathrm{d^3}N^\mathrm{AA}_\mathrm{{jet}}}{\mathrm{d}p^{\mathrm{ch}}_\mathrm{T,jet} \mathrm{d}\Delta\varphi \mathrm{d}\eta_\mathrm{jet}} \bigg|_{p_\mathrm{T,trig} \in \mathrm{TT_{Sig}}} - c_\mathrm{ref} \cdot \frac{1}{N^\mathrm{AA}_\mathrm{trig}} \frac{\mathrm{d^3}N^\mathrm{AA}_\mathrm{{jet}}}{\mathrm{d}p^{\mathrm{ch}}_\mathrm{T,jet} \mathrm{d}\Delta\varphi \mathrm{d}\eta_\mathrm{jet}} \bigg|_{p_\mathrm{T,trig} \in \mathrm{TT_{Ref}}} ,
\label{eq:DeltaRecoil}
\end{equation}


\noindent where $c_\mathrm{ref}$ accounts for the combined effects of invariance of total jet yield with $\pttrig$. $c_\mathrm{ref}$ is calculated bin-by-bin in $\Delta\varphi$, and ranges from 0.94 in
the $\Delta\varphi$ region closest to $\frac{1}{2} \pi$ to 0.86 in the $\Delta\varphi$ region closest to $\pi$.
With the $\Delta_\mathrm{recoil} $ observable one removes entirely the background from uncorrelated reconstructed jets, giving the possibility of extending jet measurements to low-$\pt$.
The jet population measured with this technique is likewise not biased in terms of jet fragmentation pattern. The $\pttrig$ trigger-track classes chosen for this analysis are $5 < \pttrig < 7~\gevc$ for the Reference class and $20 < \pttrig < 50~\gevc$ for the Signal class.

\section{Analysis}

This analysis was carried out using central 0--10\% Pb--Pb collisions at $\sqrtsNN = 5.02~\tev$ collected by ALICE in 2018. The ALICE experimental setup is detailed in Ref.~\cite{Aamodt:2008zz}. The events were triggered using both minimum bias and central triggers based on signals in the ALICE V0 detector, and further offline selection was applied to ensure background events (e.g. beam-gas events) were removed. 92M events were selected for analysis.
The measurement uses jets reconstructed from charged tracks from the ALICE Inner Tracking System and Time Projection Chamber, i.e. `track-based jets'.
To ensure statistical independence of the Signal and Reference recoil jet distributions, each event is randomly assigned to one of the TT classes, and the statistical reach of the distributions is optimised by using $80\%$ of events for the Signal subset and $20\%$ for the Reference subset.
Jet finding was performed with the FastJet package~\cite{Cacciari:2011ma} using the anti-$k_\mathrm{T}$ algorithm ~\cite{Cacciari:2008gp} with jet resolution parameter $R=0.2$, using tracks with a low-$\pt^\mathrm{ch}$ constituent cut-off of $\pt^\mathrm{ch} = 0.15~\gevc$. The underlying event density $\rho$ was then subtracted from the raw \pt ($p_{\mathrm{T,jet}}^{\mathrm{raw,ch}}$) of each jet with area $A_\mathrm{jet}$ as $p_{\mathrm{T,jet}}^{\mathrm{reco,ch}} = p_{\mathrm{T,jet}}^{\mathrm{raw,ch}} - \rho A_{\mathrm{jet}}$. 
$\rho$ was estimated event-by-event by reconstructing jets with the $k_\mathrm{T}$ algorithm and using the median \pt-density of these jets, excluding the two leading jets in the event from the median.
Events with high-$\pt$ tracks have on average a higher $\rho$ using this method, and to account for this offset, the difference between the mean $\rho$ in Signal- and Reference-classified events is added to $\rho$ in reference-classified events.

\begin{figure}[hbt]
	\centering
	\includegraphics[width=.53\textwidth]{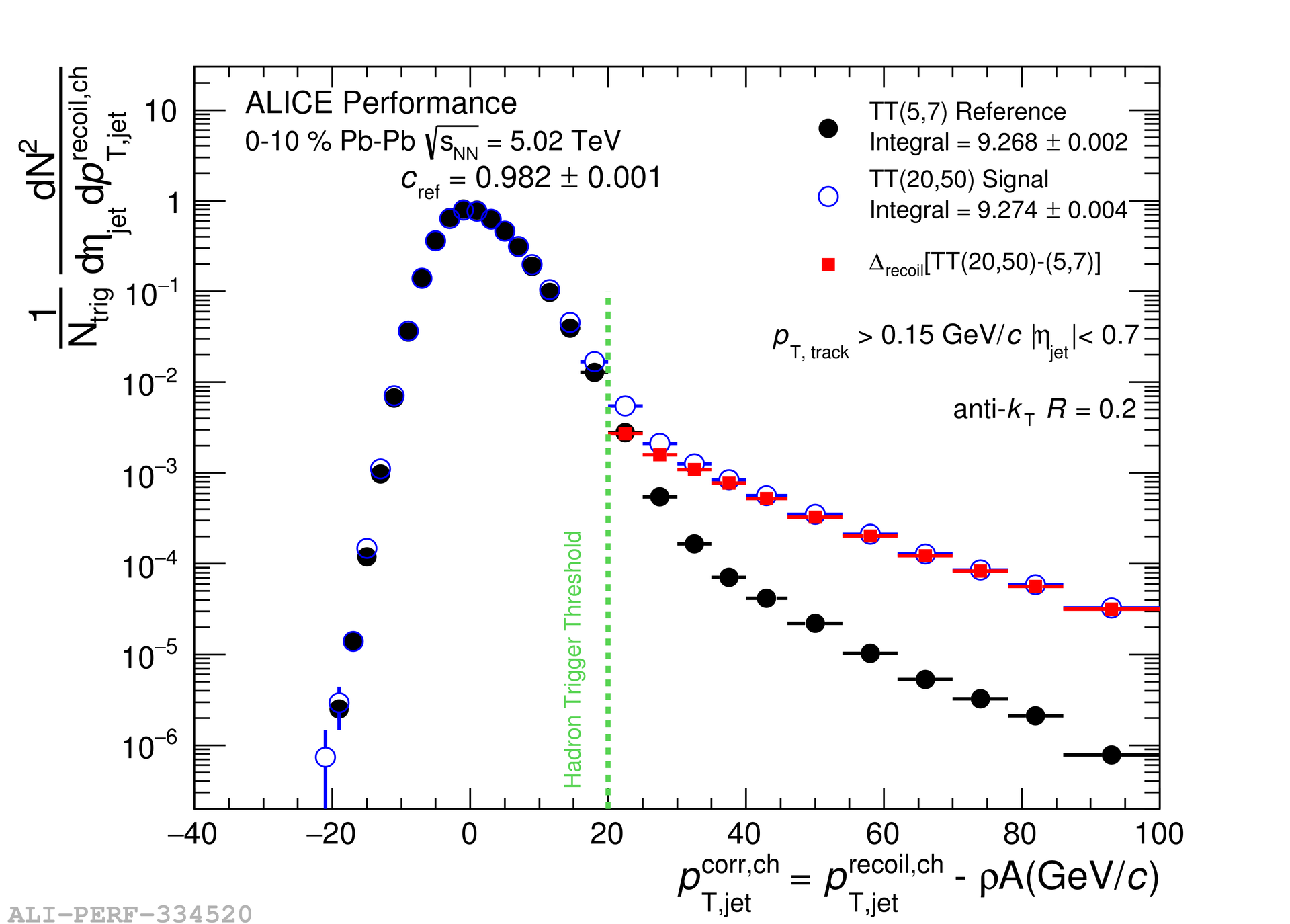}
	\includegraphics[width=.45\textwidth]{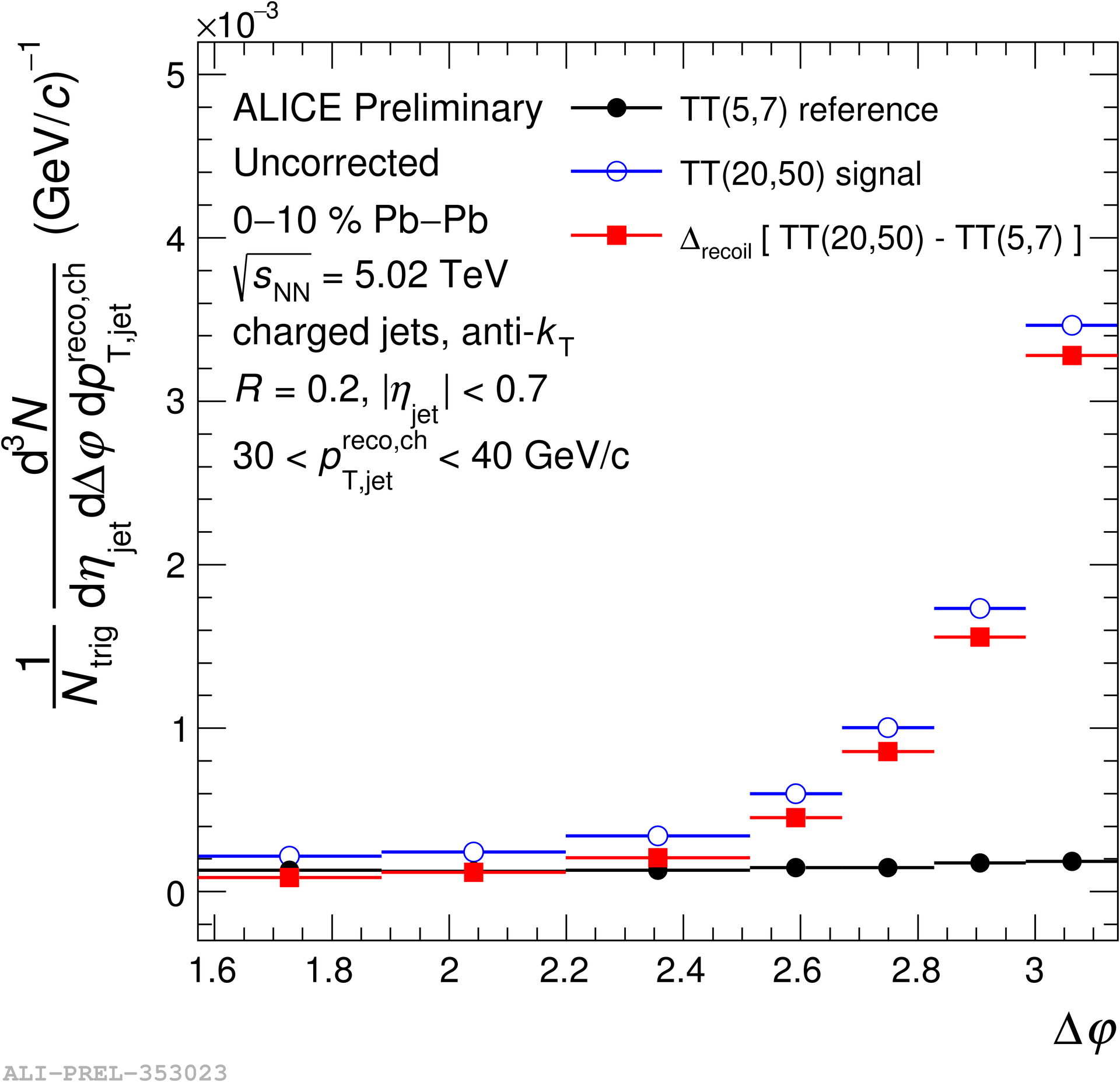}
	\caption[]{The raw trigger-normalised recoil jet distributions as a function of $\ptreco$ (left) and $\Delta\varphi$ (right).}  
	\label{fig:raw}
\end{figure}

2-dimensional distributions ($\ptreco$, $\Delta\varphi$) were filled separately for $\TTsig$ and $\TTref$ recoil jets. Figure \ref{fig:raw} shows the projection of these distributions on the $\ptreco$ and $\Delta\varphi$ axes. The $\TTsig$ and $\TTref$ distributions are seen to be similar around $\ptreco=0$  and thus are independent of $\pttrig$, corresponding to the fact that the jet population in this region is comprised primarily of combinatorial jets. At higher $\ptreco$ the distributions begin to separate. The $\Delta\varphi$ distribution in the region $30 < \ptreco < 40~\gevc$ shows a clear peak at $\Delta\varphi \sim \pi$, indicating the back-to-back topology of di-jet events.

The raw distributions were corrected simultaneously in $\ptreco$ and $\Delta\varphi$ for detector effects and residual background fluctuations using 2-dimensional Bayesian unfolding techniques~\cite{Adye:1349242}, with a response mapping detector-level $\ptreco$ and $\Delta\varphi$ to particle level. The response was constructed using simulated PYTHIA events, which were reconstructed using a full GEANT simulation, and embedded into heavy-ion data events. PYTHIA detector-level jets were reconstructed among the heavy-ion background (hybrid-level jets) and matched with the PYTHIA detector-level jets, which were then matched with PYTHIA particle-level jets, and the response was generated with matched hybrid-level and particle-level jets. 
The systematic uncertainties considered in this analysis include the uncertainties due to the tracking efficiency, the unfolding (including the choice of prior, binning and Bayesian regularisation parameter), the $c_\mathrm{ref}$ scaling, and jet matching criteria.

\section{Results}

\begin{figure}[hbt]
	\centering
	\includegraphics[width=.60\textwidth]{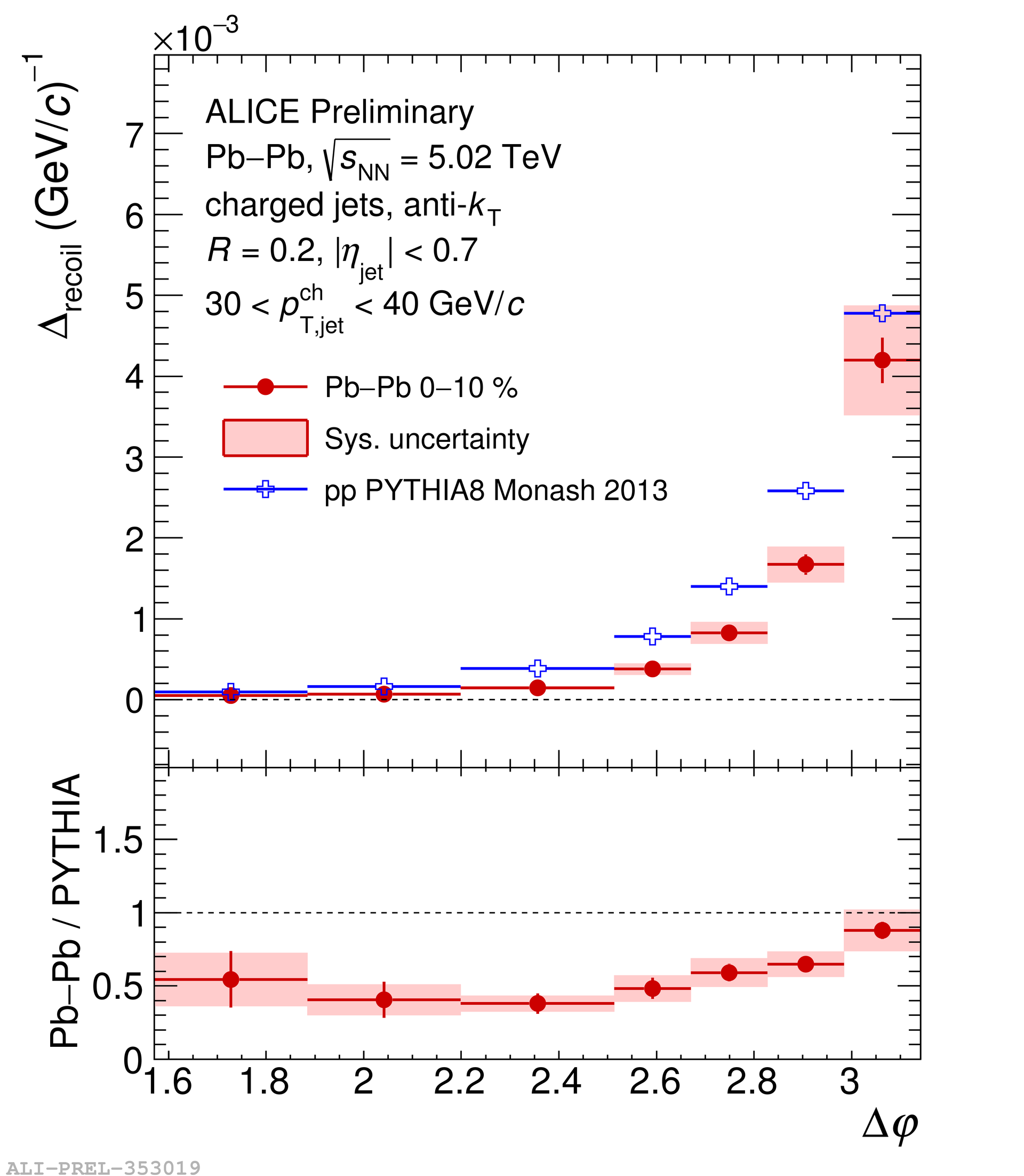}
	\caption[]{The fully corrected $\Delta_\mathrm{recoil}$ distributions in 0--10\% Pb--Pb collisions as a function of $\Delta\varphi$, compared to a PYTHIA reference.}  
	\label{fig:results}
\end{figure}

The fully-corrected $\Delta_\mathrm{recoil}$ distribution as a function of $\Delta\varphi$ in the region $30 < \ptch < 40~\gevc$ is shown in Fig. \ref{fig:results}. The reference distribution for pp collisions was generated by PYTHIA8~\cite{Sjostrand:2007gs} (Monash 2013 tune~\cite{Skands:2014pea}). The recoil jet yield in Pb--Pb collisions is observed to be suppressed with respect to the pp reference, indicating jet quenching effects. The ratio of the two distributions in the lower panel shows a clear indication that the $\Delta\varphi$ distribution is narrower in central Pb--Pb collisions, with the ratio of Pb--Pb / PYTHIA equal to around 0.9 at $\Delta\varphi \sim \pi$, and around 0.4--0.5 in the region $ \pi / 2 < \Delta\varphi < 3\pi / 4$.

\section{Outlook}

The first measurement of the fully-corrected semi-inclusive distribution of charged jets recoiling from a trigger hadron in 0--10\% Pb--Pb collisions at $\sqrt{s_{\mathrm{NN}}}$ = 5.02 TeV as a function of the azimuthal angle $\Delta\varphi$ has been shown. The results indicate a suppression and narrowing of the recoil jet yield in $30 < \ptch < 40~\gevc$ with respect to PYTHIA. For a robust interpretation of this result, it is crucial that a measurement in pp collisions is taken as a true vacuum reference. A measurement using the high statistics pp dataset at 5 TeV taken in 2017 will provide this necessary reference. Work to extend this measurement to higher jet $R$ and a wider $\ptch$ is also ongoing.

\bibliographystyle{JHEP}
\bibliography{include/bibfile}

\end{document}